\documentclass[aps,prb,twocolumn,superscriptaddress,showpacs]{revtex4}
\usepackage{amsfonts}
\usepackage{amssymb}

\begin{document}


\title{Spin correlations in rare-earth paramagnetic systems; 
neutron linewidths and $\mu$SR
spin-lattice relaxation rates.}



\author {A.~Yaouanc}
\affiliation{CEA/DSM/Institut Nanosciences et Cryog\'enie, 38054 Grenoble, France}




\date{\today}

\begin{abstract}
We consider the spin correlation functions for rare-earth compounds in their
paramagnetic state with the aim to understand the temperature dependence 
of the quasi-elastic and crystal-field linewidths and the spin-lattice relaxation
rates measured by the muon-spin-relaxation technique. Both the conduction 
electron and phonon relaxation mechanisms are treated. First the single-ion
dynamics is described using an iterative method introduced 
by P. M. Richards. Then the case of a regular lattice of rare-earth ions
is studied using the random-phase approximation (RPA). Applications to simple
crystal-field level schemes are given with particular emphasis on the phonon
relaxation mechanism. This allows us to investigate the domain of validity of 
our results. In order to account for data recorded on insulators, a
phenomelogical modification is suggested. 
   
\end{abstract}


\maketitle


The study of the spin dynamics of a regular lattice of rare-earth
ions in the paramagnetic phase of an insulator is not a subject of much 
interest in itself. However, exotic dynamics may has been
seen at low temperature in geometrically frustrated magnetic 
rare-earth compounds, see Ref.~[\onlinecite{Gardner99}] for an example 
and Ref.~[\onlinecite{Moessner06}] for a general reference.
Therefore we find useful to establish the expected thermal behaviour 
of the crystal-field linewidth and the positive muon ($\mu$SR) 
spin-lattice relaxation rate arising from the phonon-induced relaxation.

A theory for the linewidth of crystal-field excitations in metallic
rare-earth systems has been given a long time ago \cite{Becker77}. 
The description of the single-ion dynamics
was based on the memory function technique. Later on, the same problem 
was considered by Richards using a more conventional iterative 
method akin to the Redfields's, the derived system of 
equations being solved by Laplace transform \cite{Richards84}.

Here we examine the contribution of the conduction electrons and phonons 
to the spin dynamics in the paramagnetic phase of compounds with a rare-earth
sublattice. The crystal electric field acting on the magnetic ions will 
be dealt with. The interaction between the ions will be accounted for.

Two experimental methods are commonly used to investigate spin 
dynamics in magnetic materials: the neutron scattering and $\mu$SR 
techniques. While the relationship between the measured 
intensity and the physical properties of the spin correlation 
functions is well documented for the former, see for example 
Refs.~[\onlinecite{Lovesey86a,Jensen91}], it is not the case for 
the latter, at least for a lattice of rare-earth ions for which crystal-field
effects are expected to be important.
This is done in this paper, building on previous works
dealing with the critical spin dynamics and magnon-induced 
relaxation as probed by $\mu$SR, see Ref.~[\onlinecite{Dalmas04}] 
and references therein. 

The organization of this paper is as follows.  In Sec.~\ref{symmetry},
we investigate some basic properties of the spin-correlation tensor. 
In Sec.~\ref{Lambda} we write the $\mu$SR spin-lattice relaxation 
rate, $\lambda_Z$, in terms of spin-correlation 
functions. In Sec.~\ref{Correlation} these functions 
are described for the case of a single rare-earth ion,
following the method of Richards \cite{Richards84}.
In the following section, Sec.~\ref{relaxation}, we give the 
expressions for the conduction electrons and phonons correlation 
functions. The theory develops in the previous sections is 
applied for simple crystal-field level schemes in Sec.~\ref{Model}.
The interaction between the magnetic ions is taken into account 
in Sec.~\ref{Lattice}. We start by using the random-phase 
approximation (RPA). Since this approximation is found to breakdown
for a limit of experimental interest, a phenomelogical modification is proposed.

\section{Properties of the spin-correlation tensor}
\label{symmetry}

In this section, we study some basic properties, mostly 
resulting from the point symmetry at the rare-earth site, of the symmetrised
spin-correlation 
function $\Lambda_{\alpha \beta}(\omega) = \int_{- \infty}^\infty 
\left \langle \{ J_\alpha (t) J_\beta  \} \right \rangle \exp(i \omega t){\rm d} t$,
where $2 \left \langle\{ J_\alpha (t) J_\beta  \} \right \rangle
= \left \langle  J_\alpha (t) J_\beta   \right \rangle +
\left \langle  J_\beta J_\alpha (t)   \right \rangle $. $J_\alpha$ is the $\alpha$ 
component of the total angular momentum of a magnetic ion in the compound
of interest. We explicitly assume $J$ to be a good quantum number. 
This is true for a rare-earth ion.

Our first task is to examine the structure of ${\bar {\bar \Lambda}}(\omega)$,
the tensor elements of which are the functions $\Lambda_{\alpha \beta}(\omega)$.
This tensor is written in the reference frame $\{{\bf x},{\bf y},{\bf z} \}$ 
attached to the crystal lattice. In fact, because the crystal electric field potential 
is expressed in terms of Stevens operators which are polynomials 
of $J_+ = J_x + i J_y$, $J_-= J_x - i J_y$ and $J_z$, we shall work with the 
reference frame  $\{{\bf +},{\bf -},{\bf z} \}$. It is only in this representation 
that the Stevens operators display nicely their symmetry properties.

In this study we shall split the total Hamiltonian of the system under study 
into two parts. We write ${\mathcal H} = {\mathcal H}_0 + {\mathcal H}_1$ 
where ${\mathcal H}_1$ describes the perturbation of the system relative 
to ${\mathcal H}_0$. In addition, we take ${\mathcal H}_0$ as the sum of two
commutating operators, 
${\mathcal H}_0 = {\mathcal H}_{\rm CF} + {\mathcal H}_{\rm L}$, where 
the first term is modelling the crystal-field energy levels of the rare-earth 
ions and the second the lattice. We restrict ourselves to the analysis of
zero-field measurements. We shall assume in this paper 
$\left [{\mathcal H}_{\rm L},  J_\alpha (t)\right ]=0$.

Physically the relaxation of a spin correlation function of a single 
rare-earth ion is driven by the crystal-field transitions induced by the 
coupling of the crystal field with the lattice. Neglecting this 
coupling, {\sl i.e.} assuming ${\mathcal H}_1 =0$,
we determine which spin correlations are compatible with the point 
symmetry at the rare-earth site. If a correlation vanishes when the 
coupling is neglected, it cannot be large even if it is taken into 
account. Hence, we study the symmetry properties of 
$\left \langle J_\alpha (t) J_\beta \right \rangle_0$, where
\begin{eqnarray}
& & \left \langle J_\alpha (t) J_\beta  \right \rangle_0 \cr
& = &   
{\rm Tr} \left \{ \rho_0
\exp \left (i {\mathcal H}_0 t /\hbar  \right )  J_\alpha
\exp \left (-i {\mathcal H}_0 t /\hbar  \right ) J_\beta \right \}. 
\label{symmetry_1}
\end{eqnarray}
We have introduced the density operator
$\rho_0 =\exp \left (- \beta {\mathcal H}_0 \right )/Z_0 $ 
where $Z_0$ is the partition function of ${\mathcal H}_0$,
$\beta =1/(k_{\rm B} T)$ and 
${\rm Tr} \{ A \}$ stands for the trace of $A$ over the quantum states 
of ${\mathcal H}_0$. Since 
$\left [{\mathcal H}_{\rm L},  J_\alpha (t)\right ]=0$,
$\left [{\mathcal H}_{\rm L},  J_\alpha \right ]=0$ and thus
\begin{eqnarray}
& & \left \langle J_\alpha (t) J_\beta  \right \rangle_0 
 =  \left \langle J_\alpha (t) J_\beta  \right \rangle \cr  
& = &
{\rm Tr} \left \{ \rho
\exp \left (i {\mathcal H}_{\rm CF} t /\hbar  \right )  J_\alpha
\exp \left (-i {\mathcal H}_{\rm CF} t /\hbar  \right ) J_\beta \right \}. 
\label{symmetry_2}
\end{eqnarray}
$\rho$ is the density operator for the crystal-field levels.
We use the simplified notations $\rho$ and 
$\left \langle \cdots  \right \rangle$ rather than
$\rho_{\rm CF}$ and 
${\left \langle \cdots  \right \rangle}_{\rm CF}$ because there is 
no risk of confusion.

We are going to show that the only tensor elements which may not be zero 
in most physical cases are
$\Lambda_{zz}(\omega)$, $\Lambda_{+-}(\omega)$ and 
$\Lambda_{-+}(\omega)$. To proof, for example, that  $\Lambda_{-z}(\omega)$ is 
usually incompatible with the point symmetry at the rare-earth site, 
we need to show $\left \langle J_- (t) J_z \right \rangle = 0$. We have
\begin{eqnarray}
& & \left \langle J_- (t) J_z   \right \rangle 
\cr
& = &
{\rm Tr} \left \{ \rho
\exp \left (i {\mathcal H}_{\rm CF} t /\hbar  \right )  J_-
\exp \left (-i {\mathcal H}_{\rm CF} t /\hbar  \right ) J_z \right \}
\cr
& = &
{\rm Tr} \left \{ \rho
\exp \left (i {\mathcal H}_{\rm CF} t /\hbar  \right ) {\mathcal I} J_-{\mathcal I}
\exp \left (-i {\mathcal H}_{\rm CF} t /\hbar  \right ) J_z \right \}.
\label{symmetry_3}
\end{eqnarray}
We have inserted
the unit tensor ${\mathcal I}$ twice. ${\mathcal H}_{\rm CF}$ is invariant under the 
operations of the point symmetry group at the ion site.  
Let us assume that 
one of the symmetry operation of the group is a rotation of angle $\varrho= 2 \pi/n$ 
about the $z$ axis, {\sl i.e.} the $z$ axis is a n-fold symmetry axis.
Mathematically this invariance means  
\begin{eqnarray}
[ {\mathcal H}_{\rm CF} , {\mathcal R}_z (\varrho)   ]=0,
\label{symmetry_4}
\end{eqnarray}
where ${\mathcal R}_z (\varrho)= \exp(-i \varrho J_z)$ stands for the rotation operator about 
the $z$ axis of angle $\varrho$; see for example Ref.~[\onlinecite{Messiah65}].  
Recalling the relation
\begin{eqnarray}
{\mathcal R}_z (\varrho) J_- {\mathcal R}_z^{-1}(\varrho) =
\exp(-i \varrho) J_-,
\label{symmetry_5}
\end{eqnarray}
and Eq.~\ref{symmetry_4}, starting from Eq.~\ref{symmetry_3} 
we derive
\begin{eqnarray}
\left \langle J_- (t) J_z   \right \rangle 
& = & \exp(-i \varrho)
\left \langle J_- (t) J_z   \right \rangle.
\label{symmetry_6}
\end{eqnarray}
We have used 
${\mathcal I} = {\mathcal R}_z^{-1}(\varrho){\mathcal R}_z (\varrho) $.
The only solution of this equation with $\varrho =  2 \pi/n$ and $n \ge 2$ is 
$\left \langle J_- (t) J_z   \right \rangle = 0$. 
In the same way we find $\left \langle J_z  J_- (t)\right \rangle =0$. Our conclusion is 
that the existence of the symmetrised correlation function  
$\left \langle\{ J_- (t) J_z  \} \right \rangle$ 
is inconsistent with the $z$ axis being two-fold or of higher symmetry. 
Using the same procedure we find five other elements of 
${\bar {\bar \Lambda}}(\omega)$ to be zero when $n \geq 3$,
because $\left \langle\{ J_- (t) J_-  \} \right \rangle =0$ 
only if $n \geq 3$. However, this tensor is not diagonal in 
$\{{\bf +},{\bf -},{\bf z} \}$ since the three non vanishing functions are 
$\Lambda_{+-}(\omega)$,
$\Lambda_{-+}(\omega)$ and $\Lambda_{zz}(\omega)$. 

Before diagonalizing ${\bar {\bar \Lambda}}(\omega)$, we first establish some general 
properties of the symmetrised correlation functions.
As $\langle J_\alpha (t) J_\beta  \rangle^* = 
\langle J_\beta^\dagger (-t) J_\alpha^\dagger  \rangle$
and $\langle J_\alpha J_\beta (t) \rangle^* = 
\langle J_\beta^\dagger  J_\alpha^\dagger (-t)  \rangle$,
from the definition of $ \Lambda_{zz}\left (\omega  \right )$,
$ \Lambda_{+-}\left (\omega  \right )$ and  
$ \Lambda_{-+}\left (\omega  \right )$ we infer these functions to be real. In our 
notations $A^\dagger$ is the adjoint of operator $A$. Since in addition
$\langle J_\alpha (t) J_\beta  \rangle^* = 
\langle J_\beta^\dagger  J_\alpha^\dagger (t)  \rangle$
and $\langle J_\alpha J_\beta (t) \rangle^* = 
\langle J_\beta^\dagger(t)  J_\alpha^\dagger   \rangle$,
we derive $ \Lambda_{+-}\left (\omega  \right ) = \Lambda_{-+}\left (-\omega  \right )$. 
Therefore the present discussion leads to restrict the number 
of correlation functions to be computed 
to two: $ \Lambda_{zz}\left (\omega  \right ) $ and, {\sl e.g.} 
$ \Lambda_{+-}\left (\omega  \right )$. 

${\bar {\bar \Lambda}}(\omega)$ is diagonal in a frame denotes as 
$\{{\bf +}^\prime,{\bf -}^\prime,{\bf z} \}$. 
We shall need to specify explicitly only the eigenvalues. They are  
$\Lambda_{+^\prime}(\omega) = \Lambda_{+^\prime +^\prime}(\omega) = 
\sqrt{\Lambda_{+-}(\omega)\Lambda_{+-}(- \omega)}$,
$\Lambda_{-^\prime}(\omega) = \Lambda_{-^\prime -^\prime}(\omega) = -\Lambda_{+^\prime}(\omega)$
and $\Lambda_{z}(\omega) =\Lambda_{zz}(\omega)$. One easily shows that 
$\Lambda_{+-}(\omega)\Lambda_{+-}(- \omega) \geq 0$.

Let us introduce the auxilary function
\begin{eqnarray}
\Omega_{\alpha \beta} (\omega ) = \int_{-\infty}^\infty 
\left \langle J_\alpha (t) J_\beta   \right \rangle \exp(i \omega t) {\rm d} t. 
\label{symmetry_8}
\end{eqnarray}
The condition of detailed balance gives the 
relationship \cite{Lovesey86a,Jensen91}
\begin{eqnarray}
\Omega_{\alpha \beta} (\omega ) = 
\exp\left(\beta \hbar \omega  \right )  \Omega_{\beta\alpha} (-\omega ).
\label{symmetry_9}
\end{eqnarray}
This implies  
\begin{eqnarray}
\Lambda_{+-}\left (\omega  \right ) &  = &  
[1 + \exp\left (-\beta \hbar \omega  \right )] 
\Omega_{+-}\left (\omega  \right )/2,\cr
\Lambda_{zz}\left (\omega  \right ) & = & 
[1 + \exp\left (-\beta \hbar \omega \right )] 
\Omega_{zz}\left (\omega  \right )/2.
\label{symmetry_10}
\end{eqnarray}
Hence, once 
$\Omega_{zz}\left (\omega  \right )$ and $\Omega_{+-}\left (\omega  \right )$ 
are computed, $\Lambda_{+-}\left (\omega  \right ) $ and 
$\Lambda_{zz}\left (\omega  \right )$ can be evaluated.

Note that the inelastic neutron scattered intensity is directly
proportional to $\Omega_{\alpha \beta} (\omega )$; see for example
Refs.~[\onlinecite{Lovesey86a,Jensen91}]. The so-called scattering function
$S_{\alpha \beta} (\omega )$ is within a constant of proportionality equal to
$\Omega_{\alpha \beta} (\omega )$.

\section{$\mu$SR spin-lattice relaxation rate and spin-correlation functions}
\label{Lambda}

The $\mu$SR techniques are based on the detection of the positron 
produced during the muon decay. Of particular interest is the 
longitudinal field technique for which the initial muon beam polarization,
the external field ${\bf B}_{\rm ext}$ if any, and the axis of the positron 
detectors are all parallel to the same axis denoted as $Z$.
Therefore that axis is of special importance. We define it 
by its polar angles $\theta$ and $\varphi$ in the reference
frame $\{{\bf x},{\bf y},{\bf z} \}$ introduced in the previous section.
We note that the $\mu$SR techniques probe only the magnetic field
at the muon site and therefore the magnetic field correlation tensor 
${\bar {\bar \Phi}} \left (\omega_\mu  \right )$
at an energy $\hbar\omega_\mu$. $\omega_\mu$ is proportional to 
the local field at the muon site. Since the difference between that 
field and $B_{\rm ext}$ is usually very 
small \cite{Schenck95}, we write 
$\omega_\mu = \gamma_\mu B_{\rm ext}$ where $\gamma_\mu$
is the muon gyromagnetic ratio 
($\gamma_\mu$ = 851.6 Mrad~s$^{-1}$~T$^{-1}$).
${\bar {\bar \Phi}} \left (\omega_\mu  \right )$ is also defined in 
$\{{\bf +}^\prime,{\bf -}^\prime,{\bf z}^\prime \}$.

A longitudinal field $\mu$SR experiment consists of measuring the 
relaxation function which is, to a good approximation, an exponential function 
in the case of interest here, {\sl e.g.} for paramagnets with a regular
lattice of rare-earth ions, and 
therefore is characterized by a single parameter, the spin-lattice 
relaxation rate $\lambda_Z$. Implicit is the assumption that the motional
narrowing limit applies, {\sl i.e.} the fluctuations of the rare-earth moments 
are rapid. $\lambda_Z$ is directly related to  
${\bar {\bar \Phi}} \left (\omega_\mu  \right )$:
\begin{eqnarray}
\lambda_Z & = & {\gamma_\mu^2  \over 2} \sum_{\gamma,\beta} 
L_{\gamma \beta } (\theta, \varphi)
\Phi_{\beta \gamma} \left (\omega_\mu  \right ),
\label{Lambda_1}
\end{eqnarray}
where now $\{ \gamma, \beta \} = \{+^\prime,-^\prime,z\}$, the matrix elements 
$L_{\gamma \beta} (\theta, \varphi)$ can be computed from the material given in 
Appendix C of Ref.~[\onlinecite{Dalmas97}]
and $\Phi_{\beta \gamma} \left (\omega_\mu  \right )$
is an element of ${\bar {\bar \Phi}} \left (\omega_\mu  \right )$. 
We have introduced the Fourier transform 
$\Phi_{\beta \gamma} \left (\omega  \right ) = \int_{- \infty}^\infty 
\Phi_{\beta \gamma} \left ( t \right ) \exp(i \omega t){\rm d} t$, 
a definition different from the one used previously \cite{Dalmas96}. 
The total angular momenta of the 
magnetic ions interact with the muon spin through the dipole 
interaction and, in addition, the hyperfine interaction which 
accounts for the transfered magnetic field for an insulator and 
the conduction electrons for a metallic compound. This induces a 
magnetic field at the muon site
with correlation functions such that
\begin{eqnarray}
\Phi_{\beta \gamma} \left (\omega  \right ) &=  & 
\left ({\mu_0  \over 4 \pi}  \right )^2 g^2 \mu_{\rm B}^2
{ 1 \over v^2}\sum_{\alpha} \sum_{i, j}
G_{{\bf r}_i}^{\beta \alpha} G_{{\bf r}_j}^{\alpha \gamma}
\Lambda_{\alpha;{i,j}}\left (\omega  \right ). \cr
& &
\label{Lambda_2}
\end{eqnarray}
$i$ and $j$ run over the magnetic ions. 
$\Lambda_{\alpha;{i,j}}\left (\omega  \right )$
is a spin-correlation function between sites $i$ and $j$. $G_{{\bf r}_i}$ 
is the coupling tensor labeled by the vector ${\bf r}_i$ which points 
to site $i$ and the origin of which is the muon site. Explicitly
we have
\begin{eqnarray}
G_{{\bf r}_i}^{\beta \alpha} & = & v \left( {3 r_{i,\beta} r_{i,\alpha}\over r_i^5} -
{\delta^{\beta \alpha }  \over r_i^3}   \right)
+ H_{r_i}\delta^{\beta \alpha }. 
\label{Lambda_3}
\end{eqnarray}
$H_{r_i}$ is the hyperfine constant for the coupling to the ion $i$. 
For simplicity we assume all the magnetic ions to be equivalent: they
sit on a Bravais lattice.
$\mu_0$ is the magnetic permeability of free space, $g$ the spectroscopic 
factor, $\mu_{\rm B}$ the electronic Bohr magneton
and $v$ the volume per magnetic ion. 

Two approaches are now possible. If the interaction between 
the total angular momenta of the magnetic ions is negligible, 
we can restrict ourselves to the behaviour of a single momentum. 
Otherwise their interactions has to be taken into account.

We first consider the case for which 
$\Lambda_{\alpha;{i,j}}\left (\omega  \right )=
\Lambda_{\alpha}\left (\omega  \right ) \delta_{i,j}$. The intersite
correlations are therefore negligible and thus
\begin{eqnarray}
\Phi_{\beta \gamma} \left (\omega  \right ) & = & 
\left ({\mu_0  \over 4 \pi}  \right )^2 {g^2 \mu_{\rm B}^2
\over v^2}\sum_{\alpha} \Lambda_{\alpha}\left (\omega  \right )
\sum_{i}
G_{{\bf r}_i}^{\beta \alpha } G_{{\bf r}_i}^{\alpha \gamma }.\cr
& &
\label{Lambda_4}
\end{eqnarray}
Hence, based on our discussion in Sec.~\ref{symmetry},
$\lambda_Z$ is found to be the weighted sum of two spin-correlation functions
for a compound with at least a three-fold symmetry axis: they are
$\Lambda_{z}\left (\omega  \right ) =\Lambda_{zz}\left (\omega  \right )$
and, {\sl e.g.} $\Lambda_{+^\prime}(\omega) = 
\sqrt{\Lambda_{+-}(\omega)\Lambda_{+-}(- \omega)}$. Obviously, additional symmetry 
elements of the point group at the rare-earth site may reduce the number of  
correlations to be computed to only one.

It is convenient to use the space-Fourier transform 
$\Lambda_{\alpha}({\bf q},\omega)$ if the intersite correlations are appreciable.
Referring to Ref.~[\onlinecite{Yaouanc93}],
\begin{eqnarray}
& & \Phi_{\beta \gamma} \left (\omega  \right ) \cr
& = & 
\left ({\mu_0  \over 4 \pi}  \right )^2 {g^2 \mu_{\rm B}^2
\over V }\sum_{\alpha} \int
{\mathcal G}^{\beta \gamma \alpha} ({\bf q}) 
\Lambda_{\alpha} \left ({\bf q},\omega  \right ) 
{{\rm d}^3 {\bf q}  \over (2 \pi )^3}, \cr
& & 
\label{Lambda_5}
\end{eqnarray}
with the definitions ${\mathcal G}^{\beta \gamma \alpha} ({\bf q}) =
G^{\beta \alpha} ({\bf q}) G^{\alpha \gamma}(- {\bf q})$ and
$G^{\beta \alpha} ({\bf q}) = \sum_i G_{{\bf r}_i}^{\beta \alpha }
\exp \left (i {\bf q} \cdot {\bf r}_i   \right ).$
$V$ is the volume of the sample, {\sl i.e.} $V = N v$ where $N$ is the number
of magnetic ions in the sample. The integration is over the Brillouin zone.
Again, two spin-correlation functions are sufficient to model $\lambda_Z$
for a compound with at least a three-fold symmetry axis. As pointed out above, the number
may be reduce to only one due to the presence of additional symmetries at the rare-earth 
site.

\section{Single-ion correlation functions 
$\Omega_{\alpha \beta}\left (\omega  \right )$ in the 
reference frame $\{{\bf +},{\bf -},{\bf z} \}$}
\label{Correlation}

In this section we give formulae for $\Omega_{zz}\left (\omega  \right )$
and $\Omega_{+-}\left (\omega  \right )$. The interaction between the 
rare-earth ions is neglected. The correlation functions are evaluated with
the Richards's method \cite{Richards84}. In fact, it is the Laplace transform 
of $J_\alpha(t)$ which is considered there. P. M. Richards has only presented 
the result of his work, focusing on the electronic relaxation. 
Here we sketch the derivation and apply the result for two mechanisms which 
can drive the crystal-field transitions: the conduction electrons and 
the strain field. 

The interaction representation is used. This requires to introduce 
the two new operators, $J^*_\alpha (t)$ and ${\mathcal H}^*_1 (t)$ 
(see for example Ref~[\onlinecite{Slichter96}]), such that
\begin{eqnarray}
J^*_\alpha (t)  & = & \exp \left (-i {\mathcal H}_0 t /\hbar  \right ) 
J_\alpha (t) \exp \left (i {\mathcal H}_0 t /\hbar  \right ), \cr
{\mathcal H}^*_1 (t)  & = & \exp \left (-i {\mathcal H}_0 t /\hbar  \right ) 
{\mathcal H}_1  \exp \left (i {\mathcal H}_0 t /\hbar  \right ).
\label{Corr_1}
\end{eqnarray}
Note that $J^*_\alpha (t=0) = J_\alpha ( t=0)$. 

Using the Heisenberg evolution equation for $J_\alpha (t)$, 
\begin{eqnarray}
& & {{\rm d } J_\alpha (t) \over {\rm d} t} -
{i \over \hbar} \left [{\mathcal H}_0 ,  J_\alpha (t)\right ] \cr
& = & \exp \left (i {\mathcal H}_0 t /\hbar  \right ) 
 {{\rm d } J^*_\alpha (t) \over {\rm d} t}
\exp \left (-i {\mathcal H}_0 t /\hbar  \right ) ,
\label{Corr_2}
\end{eqnarray}
with
\begin{eqnarray}
{{\rm d } J^*_\alpha (t) \over {\rm d} t}  & = & 
{i \over \hbar} \left [{\mathcal H}^*_1 (t),  J^*_\alpha (t)\right ]
.
\label{Corr_3}
\end{eqnarray}
Solving this last equation for $J^*_\alpha (t)$, 
\begin{eqnarray}
J^*_\alpha (t)  & = & J_\alpha +
{i \over \hbar} \int_0^t \left [{\mathcal H}^*_1 (t^\prime),  
J^*_\alpha (t^\prime)\right ]{\rm d} t^\prime.
\label{Corr_4}
\end{eqnarray}
We are basically interested by the matrix elements of Eq.~\ref{Corr_2}.
For their evaluation, we introduce the expression for 
$J^*_\alpha (t)$ in the commutator
of Eq.~\ref{Corr_3}, dropping the first term on the right hand side
which is time independent.
We note the kets of ${\mathcal H}_{\rm CF}$ as 
$\{ |m \rangle, |n \rangle,  \cdots \}$ with, 
for example, ${\mathcal H}_{\rm CF}| m\rangle = \hbar \omega_m |m \rangle$. 
Denoting $\omega_{mn} = \omega_{m}- \omega_{n}$ and recalling the 
assumption $\left [{\mathcal H}_{\rm L},  J_\alpha (t)\right ]=0$, we 
derive
\begin{eqnarray}
& & \left \langle m \left |{{\rm d } J_\alpha (t) \over {\rm d} t} 
\right | n \right \rangle  - i\omega_{mn}
\left \langle m \left | J_\alpha (t)  \right | n \right \rangle \cr
& = & I + II + III + IV,
\label{Corr_5}
\end{eqnarray}
where the four terms labelled by $I, \cdots, IV$ have similar mathematical 
structures. 

We need to specify ${\mathcal H}_1$ at this junction. 
We focus first on the crystal-field fluctuations induced by conduction 
electrons. The exchange 
interaction between a rare-earth ion and the conduction electron 
can be written \cite{Jensen91} 
\begin{eqnarray}
{\mathcal H}_1 = {\mathcal H}_{\rm el} = -2 I_{\rm ex} (g-1) {\bf J} \cdot {\bf s},
\label{Corr_6}
\end{eqnarray}
where $I_{\rm ex}$ is an exchange integral, $g$ the Land\'e factor and ${\bf s}$ 
the electron spin operator. Then, for example,   
\begin{widetext}
\begin{eqnarray}
I  & = & -{1 \over \hbar^2} \left \langle m \left |
\int_0^t  {\mathcal H}_1
\exp \left [i {{\mathcal H}_0 \over \hbar} \left (t - t^\prime \right )\right ] 
 {\mathcal H}_1 J_\alpha (t^\prime)  
\exp \left [-i {{\mathcal H}_0 \over \hbar} \left (t - t^\prime \right )\right ] 
 {\rm d t^\prime } \right |n \right \rangle, \cr
& = &  -{4 I_{\rm ex}^2 (g-1)^2 \over \hbar^2}  \int_0^t \sum_{p,q}\sum_{\gamma,\gamma^\prime} 
s_\gamma\left \langle m \left |J_\gamma \right | p \right \rangle
\exp \left [i \omega_{pn}\left (t - t^\prime \right )\right ] 
s_{\gamma^\prime}\left ( t - t^\prime \right )
\left \langle p \left |J_{\gamma^\prime} \right | q \right \rangle
\left \langle q \left | J_\alpha(t^\prime)  \right |n \right \rangle {\rm d t^\prime }, \cr
&= & -  \sum_{p,q} G^{\rm el}_{mp,pq}\int_0^t
\exp \left [i \omega_{pn}\left (t - t^\prime \right )\right ] 
f_{\rm el}\left [ -\left ( t - t^\prime \right ) \right ]
\left \langle q \left | J_\alpha(t^\prime)  \right |n \right \rangle
{\rm d t^\prime },
\label{Corr_7}
\end{eqnarray}
\end{widetext}
with $G^{\rm el}_{k \ell, mn}= \left [{4 I_{\rm ex}^2 (g-1)^2 / \hbar^2}\right ]
\sum_\gamma
\left \langle k \left | J_\gamma  \right | \ell \right \rangle
\left \langle m \left | J_\gamma  \right |n \right \rangle$.
In the last step we have performed an average over the electronic states 
assuming the electronic correlation to be isotropic and given by
$f_{\rm el} (t) = {\left \langle s_x(t) s_x \right \rangle}_{\rm el}$.

Before discussing the other terms for the electronic relaxation,
we pay attention to the $I$ expression for the phonon-induced relaxation.
From numerous electron paramagnetic resonance (EPR) studies, see for example 
Ref.~[\onlinecite{Abragam70}], the phonon-induced relaxation between electronic 
energy levels is known to be driven by the modulation of the crystal field. We 
expand that field in powers of the strain and limit the expansion to the linear
term denoted as ${\mathcal H}_{\rm sp}$. It accounts 
for the spin-phonon interactions, {\sl i.e.} the magneto-elastic
coupling, and represents an additional electric 
potential generated by the lattice vibrations. To determine its form, we should
study in details the vibration-induced distorsion of the surrounding of the ion of
interest \cite{Poole71}. Here we follow a much simpler path. We take an 
average strain $\epsilon$ and ignore any directional properties. This results to
\begin{eqnarray}
{\mathcal H}_1= {\mathcal H}_{\rm sp}= \epsilon V.
\label{Corr_9}
\end{eqnarray}
To second order in ${\bf J}$, or more
practically using the Stevens operators  $O^m_2$ (these 
operators are listed in different references, {\sl e.g.}
Refs.~[\onlinecite{Orbach61,Abragam70}]), we have   
\begin{eqnarray}
V & = & g_{0}O^{0}_2 +  g_{1}O^{1}_2
 +  g_{-1}O^{-1}_2 +  
g_{2}O^{2}_2 +  g_{-2}O^{-2}_2, \cr & &  
\label{Corr_8}
\end{eqnarray}
where the five $g_{n}$ are unknown real parameters. 
Only the operators in this sum which break the point symmetry at the 
rare-earth site are effective in inducing the crystal-field 
transitions. Practically, to limit the number of parameters, one may have 
to select one or two out of the sum in a somewhat arbitrary fashion.
Focusing again on the $I$ term,
\begin{widetext}
\begin{eqnarray}
I  & = & -\sum_{p,q} G^{\rm ph}_{mp,pq}\int_0^t
\exp \left [i \omega_{pn}\left (t - t^\prime \right )\right ] 
f_{\rm ph}\left [ -\left ( t - t^\prime \right ) \right ]
\left \langle q \left | J_\alpha(t^\prime)  \right |n \right \rangle
{\rm d t^\prime },
\label{Corr_11}
\end{eqnarray}
\end{widetext}
with $G^{\rm ph}_{k \ell, mn}= 
\left \langle k \left | V  \right | \ell \right \rangle
\left \langle m \left | V  \right |n \right \rangle/\hbar^2$ and
$f_{\rm ph} (t) = 
{\left \langle \epsilon(t) \epsilon \right \rangle}_{\rm ph}$.
Here the average is done over the phonon bath.

It is obvious that the $I$ term for the electronic and phonon relaxation processes
have the same mathematical structure. This is also true for the three other terms,
that is, $II$, $III$ and $IV$. Therefore we need to solve Eq.~\ref{Corr_5} 
for $J_\alpha(t)$ with the four terms on the right hand side having the 
mathematical structure of Eq.~\ref{Corr_7} (equivalent to Eq.~\ref{Corr_11}). 
Since each term is expressed as a sum of convolution products, a 
Laplace transform technique is well suited. 
That ${\mathcal H}_1$ factorises into lattice and crystal-field potentials 
is seen to be a key ingredient which allows to use the Laplace transform technique.
With the notation
$Z_{\alpha,mn}(s) =\int_0^\infty \exp(-s t) \langle m|J_\alpha(t) | n \rangle{\rm d t }$,
\begin{widetext}
\begin{eqnarray}
\left (s -i \omega_{mn} \right )Z_{\alpha,mn}(s) & = &
 - \sum_{p,q} G_{mp,pq} 
\, h \left (is + \omega_{pn}  \right ) Z_{\alpha,qn} (s) 
 -  \sum_{p,q} G_{pq,qn}  
\, g \left (is + \omega_{mq}  \right ) Z_{\alpha,mp} (s) \cr
 & + &   \sum_{p,q} G_{mp,qn}  
\left [h \left (is + \omega_{pn} \right ) 
+ g \left (is + \omega_{mq}  \right )\right ]Z_{\alpha,pq} (s)
+ \left \langle m \left | J_\alpha \right | n \right \rangle.
\label{Corr_12}
\end{eqnarray}
\end{widetext}
We denote 
$g(\omega)= \int_0^\infty f(t) \exp(i\omega t){\rm d t }$ and 
$h(\omega)= \int_0^\infty f(-t) \exp(i\omega t){\rm d t }$. 
$G_{k \ell, mn}$ and $f(t)$ stand for  
$G^{\rm el}_{k \ell, mn}$ and $f_{\rm el}(t)$ or 
$G^{\rm ph}_{k \ell, mn}$ and $f_{\rm ph}(t)$, depending on the relaxation
mechanism under study. Setting $s = \varepsilon + i\omega$ with 
 $\varepsilon \rightarrow 0^+$ in the previous system 
of equations, 
\begin{widetext}
\begin{eqnarray}
i \left (\omega -\omega_{mn} \right )Z_{\alpha,mn} (i\omega) 
 + \sum_{p,q} G_{mp,pq} 
\, h \left (-\omega + \omega_{pn}  \right ) Z_{\alpha,qn}  (i\omega) 
 +  \sum_{p,q} G_{pq,qn} 
\, g \left (-\omega + \omega_{mq}  \right ) Z_{\alpha,mp} (i\omega) \cr
 -   \sum_{p,q} G_{mp,qn}  
\left [h \left (-\omega + \omega_{pn} \right ) 
+ g \left (-\omega + \omega_{mq}  \right )\right ]Z_{\alpha,pq}(i\omega) 
= \left \langle m \left | J_\alpha \right | n \right \rangle -
{\langle J_\alpha \rangle}\delta_{mn}.
\label{Corr_13}
\end{eqnarray}
\end{widetext}
As argued in Sec.~\ref{relaxation}, it is a good approximation to
neglect the imaginary part of $h(\omega)$ and $g(\omega)$. We note that
\begin{eqnarray}
h(\omega)= g(-\omega) =
\exp \left (- \beta \hbar \omega \right )g(\omega). 
\label{Corr_14}
\end{eqnarray}
The inclusion of the term ${\langle J_\alpha \rangle}\delta_{mn}$ 
guarantees that $\langle J_\alpha(t) \rangle$ relax toward its thermal 
equilibrium value. This is further discussed for the Redfield theory
by, for example, Slichter \cite{Slichter96}. 

In fact, as explained in Sec.~\ref{symmetry}, 
$\Omega_{+-} (\omega )$ and $\Omega_{z z} (\omega )$ are required. From their
definitions,
\begin{widetext}
\begin{eqnarray}
\Omega_{+- } (\omega ) & = & {1 \over Z_{\rm CF}}
\sum_{m,n}\exp \left (-\beta E_m \right )Z_{+,mn} (-i \omega)   
\left \langle n \left | J_-\right | m \right \rangle + 
{\rm c. c.},
\cr
\Omega_{zz } (\omega ) & = &{1 \over Z_{\rm CF}}
\sum_{m,n}\exp \left (-\beta E_m \right ) Z_{z,mn} (-i \omega)  
\left \langle n \left | J_z\right | m \right \rangle + 
{\rm c. c.}. \cr
& & 
\label{Corr_15}
\end{eqnarray}
\end{widetext}
${\rm c. c.}$ stands for the complex conjugate of the expression on the left 
of this symbol.  $Z_{\rm CF}=\sum_m \exp \left (- \beta E_m  \right )$ 
is the partition function for the crystal-field level scheme of 
a rare-earth ion and the $Z_{\alpha,mn}$ functions are obtained by solving 
the system of coupled $(2 J + 1)^2$ linear equations given at 
Eq.~\ref{Corr_13}.  

\section{Conduction electron and phonon correlation functions}
\label{relaxation}

In this section we evaluate the correlation functions for the 
conduction electrons and phonons. 

Turning our attention to the electronic correlation, we
need to compute 
\begin{eqnarray}
g_{\rm el}(\omega)= \int_0^\infty  
f_{\rm el}(t) \exp(i\omega t){\rm d t }.
\label{relaxation_1}
\end{eqnarray}
We shall denote $c^+_{{\bf k},\sigma}$ and $c_{{\bf k},\sigma}$ the creation 
and annihilation electron operators of wavevector ${\bf k}$ and
spin $\sigma$ and ${\mathcal H}_e$ the Hamiltonian of the electron bath.
The electrons are assumed to be free-like. We use the fermion representation 
of $s_x$, {\sl i.e.}
\begin{eqnarray}
s_x ={ 1 \over 2}\sum_{{\bf k}^\prime,{\bf k}}
\left (c^+_{{\bf k}^\prime,\downarrow}c_{{\bf k},\uparrow}
+ c^+_{{\bf k}^\prime,\uparrow}c_{{\bf k},\downarrow} \right ).
\label{relaxation_3}
\end{eqnarray}
Without lost of generality, the ion is taken to sit at the origin of the 
coordinates. $f_{\rm el} (t)$ reduces to the sum of two quantities which 
occur to be equal. Let us focus on one of them:
\begin{widetext}
\begin{eqnarray} 
& & {1 \over4  N^2} \sum_{{\bf k}, {\bf k^\prime}}\sum_{{\bf q}, {\bf q^\prime}}  
{\left \langle 
\exp \left (i {\mathcal H}_e t /\hbar  \right )
c^+_{{\bf k}^\prime,\uparrow}c_{{\bf k},\downarrow}
\exp \left (-i {\mathcal H}_e t /\hbar  \right )
c^+_{{\bf q}^\prime,\downarrow}c_{{\bf q},\uparrow}
\right \rangle}_{\rm el}  \cr
& = &
{1 \over 4 N^2} \sum_{{\bf k}, {\bf k^\prime}}\sum_{{\bf q}, {\bf q^\prime}}    
\exp \left [i 
\left (\omega_{{\bf k}^\prime,\uparrow} - \omega_{{\bf k},\downarrow}\right )t  \right ]
{\left \langle 
c^+_{{\bf k}^\prime,\uparrow}c_{{\bf k},\downarrow}
c^+_{{\bf q}^\prime,\downarrow}c_{{\bf q},\uparrow}
\right \rangle}_{\rm el} \cr
& = & 
{1 \over 4 } \sum_{{\bf k}, {\bf k^\prime}}\sum_{{\bf q}, {\bf q^\prime}}    
\exp \left [i 
\left (\omega_{{\bf k}^\prime,\uparrow} - \omega_{{\bf k},\downarrow}\right )t  \right ]
\delta_{{\bf k}^\prime, {\bf q}} \delta_{{\bf q}^\prime, {\bf k}}
{\left \langle 
c^+_{{\bf k}^\prime,\uparrow}c_{{\bf q},\uparrow}\right \rangle}_{\rm el}
{\left \langle c_{{\bf k},\downarrow}c^+_{{\bf q}^\prime,\downarrow}
\right \rangle}_{\rm el} 
 = 
{1 \over 4 } \sum_{{\bf k}, {\bf q}}  
\exp \left [i 
\left (\omega_{\bf q} - \omega_{\bf k}\right )t  \right ]
n_{\bf q}\left (1 - n_{\bf k} \right).
\label{relaxation_4}
\end{eqnarray}
\end{widetext}
$n_{\bf q}$ is the Fermi distribution function.   
The second line results after inserting in the first line the identity operator between 
the first two fermionic operators and using the Heisenberg equation. An Hartree-Fock 
decoupling is done and the spin dependence of the energy is neglected at the 
third line. The imaginary part of $g_{\rm el}(\omega)$ 
leads to a slight energy shift
and the real part describes the crystal-field relaxation. We shall neglect the shift 
and thus we only keep the real part of $g_{\rm el}(\omega)$.
It is probably difficult to detect experimentally the energy shift. We compute
\begin{eqnarray}
g_{\rm el}(\omega) & = &{\pi \over 2} \hbar \left [ N \left (E_{\rm F}  \right ) \right ]^2
{\hbar \omega \over 1 - \exp \left (-\beta \hbar \omega  \right )}.
\label{relaxation_5}
\end{eqnarray}
$N \left (E_{\rm F}  \right )$ is the density at the Fermi level per spin.

We now evaluate the phonon correlation function. 
It is convenient to express 
$\epsilon$ in terms of its Fourier components. Since the rare-earth ion can be 
taken at 
the origin of the coordinates (see for example Ref.~[\onlinecite{Kittel63}]),
\begin{eqnarray}
\epsilon 
& = &   i \sum_{\bf k} 
\sqrt {{ k^2 \hbar  \over 2 N M \omega_{\bf k}} } 
\left ( a_{\bf k} - a_{\bf k}^+ \right ).
\label{relaxation_8}
\end{eqnarray}
$N$ is the number of rare-earth ions in the compound and $N M$ its mass.  
$a^+_{\bf k}$ and $a_{\bf k}$ are the creation 
and annihilation phonon operators. Note that only a single branch of phonons
is accounted for. It is a simple matter to derive 
\begin{eqnarray}
 f_{\rm ph} (t)
& = &   {1 \over N} \sum_{\bf k} 
{ k^2 \hbar  \over 2 M \omega_{\bf k}} \cr
& \times &
\left [ \exp \left (-i \omega_{\bf k} t  \right ) 
\left (b_{\bf k} +1  \right) +
\exp \left (i \omega_{\bf k} t  \right )
b_{\bf k} \right ].
\label{relaxation_9}
\end{eqnarray}
$b_{\bf k}$ is the Bose distribution function.
Because of energy conservation during the scattering process,
the second term cannot contribute to the relaxation when $\omega > 0$. We compute
\begin{eqnarray}
 g_{\rm ph} (\omega)
& = &  {3 \pi \over N} \sum_{\bf k} 
{ k^2 \hbar  \over 2  M \omega_{\bf k}} \left (b_{\bf k} +1  \right)
\delta \left (\omega - \omega_{\bf k}\right ) \cr
 & = &{3 \over 4 \pi} {1 \over \varrho v_s^5 \hbar^2}
{\hbar^3 \omega^3 \over 1 - \exp \left (-\beta \hbar \omega  \right )}.
\label{relaxation_10}
\end{eqnarray}
The imaginary part of $ g_{\rm ph} (\omega)$ is again neglected. 
The factor $3$ is added to account for the three phonons branches.
The ${\bf k}$ sum, which extends over the Brillouin zone, is performed
assuming $\omega_{\bf k} = v_s k$. $\varrho$ is the crystal density 
and $v_s$ the sound velocity.

\section{Model computations of the correlation functions and the 
$\mu$SR spin-lattice relaxation rate}
\label{Model}

In this section we present analytical results obtained for simple rare-earth 
level schemes, our purpose being to understand the implications, on the 
correlation functions and $\mu$SR relaxation rate, of the 
crystal-field-induced relaxation by electrons and phonons.
Here we shall focus our analysis on $\Omega_{zz}\left (\omega \right ) $.
As mentionned in Sec.~\ref{Lambda}, $\lambda_Z$ is the weighted sum of
two spin-correlation functions for a compound with at least a three-fold 
symmetry axis. We shall assume, for simplicity, $\lambda_Z$ to be 
proportional only to $\Lambda_{zz}\left (\omega = 0 \right ) 
= \Omega_{zz}\left (\omega  =0 \right ) $.
  
We first discussed a result obtained from the iterative method 
detailed here against a published one derived from the memory function
method that was applied to the electronic relaxation of a Ce$^{3+}$ in 
cubic symmetry \cite{Becker77}. 

The six crystal-field levels of Ce$^{3+}$ split into a doublet $\Gamma_7$ and a 
quartet $\Gamma_8$ located at such a high energy than it can be neglected
for our purpose. We label the two states of $\Gamma_7$ as 
$| {\bar 1} \rangle$ and $| {\bar 2} \rangle$. They are 
$| {\bar 1} \rangle = (1/6)^{1/2}|~5/2~\rangle - (5/6)^{1/2}|~-3/2~\rangle$ and 
$| {\bar 2} \rangle = (1/6)^{1/2}|~-5/2~\rangle - (5/6)^{1/2}|~3/2~\rangle$. 
We compute $\langle {\bar 1} | J_z| {\bar 2} \rangle =
\langle {\bar 2} | J_z| {\bar 1} \rangle =
0$ and $\langle {\bar 2} | J_z| {\bar 2} \rangle = -
\langle {\bar 1} | J_z| {\bar 1} \rangle $. It is easily shown that
$Z_{z,12} = Z_{z,21} = 0$ and $Z_{z,11} = - Z_{z,22}$.
Therefore we only need to study Eq.~\ref{Corr_13} for 
$m=n=1$. We expect $\Omega_{zz}\left (\omega \right ) $
to be a Lorentzian function. This is obtained setting $\omega =0$
in the arguments of the $h$ and $g$ functions. This low-energy 
approximation will be always done in this 
section for simplicity. Obviously, it can be easily overcome. We derive: 
\begin{eqnarray}
i \, \omega Z_{z,11} + 4 h_{\rm el} (\omega =0) G^{\rm el}_{12,21}Z_{z,11} 
=  \langle {\bar 1} | J_z| {\bar 1} \rangle .
\label{Model_1}
\end{eqnarray}
The relaxation arises from a $G_{k \ell, mn}$
term that satisfies, as expected, the secular condition 
$\omega_{k \ell} =\omega_{m n}$ \cite{Slichter96}. 
We compute
\begin{eqnarray}
\Omega_{zz}(\omega) = {g_{\Gamma_7}^2  \over 2} 
{\Gamma_{z,{\rm el}} \over \omega^2 + \Gamma_{z,{\rm el}}^2},
\label{Model_2}
\end{eqnarray}
with the Korringa law \cite{Korringa50}
\begin{eqnarray} 
\Gamma_{z,{\rm el}} = 4 \pi g_{\Gamma_7}^2 
\left [I_{\rm ex} (g-1) N(E_{\rm F})  \right ]^2
{k_{\rm B}T \over \hbar}.
\label{Model_3}
\end{eqnarray}
$g_{\Gamma_7}= 2 \langle {\bar 2}| J_z | {\bar 2} \rangle = 5/3$ is the 
spectroscopic factor of $\Gamma_7$. This result was published by Becker 
and collaborators \cite{Becker77}. From a straightforward computation,
\begin{eqnarray} 
\Lambda_{zz}(\omega =0) = {1 \over  8 \pi} 
{\hbar \over \left [I_{\rm ex} (g-1) N(E_{\rm F})  \right ]^2
k_{\rm B}T}.
\label{Model_4}
\end{eqnarray}
This model predicts $\lambda_Z \propto 1/T$. Such a behaviour was first 
reported for ErAl$_2$ \cite{Hartmann86}.
According to Eq.~\ref{Model_1}, the linewidth is proportional 
to $h_{\rm el}(\omega=0)$. If we had considered the phonon relaxation mechanism, we 
would have found the linewidth to be proportional 
to $h_{\rm ph}(\omega=0)$. Hence,  
$\Gamma_{z,{\rm ph}} =0$, {\sl i. e.} in contrast to the electronic case 
the coupling of a doublet to the strain field does not broaden 
the quasi-static scattering.

The work reported in this paper has been originally motivated
by the observation for some geometrically frustrated magnetic materials
of a strong temperature dependence of $\lambda_Z$ at high 
temperature and sometimes a plateau at low temperature for measurements 
performed in zero field \cite{Gardner99,Dalmas03,Dalmas04a,Lago05}.
These materials are insulators, and therefore only the phonon
mechanism can play a role. 

The physics at high temperature can be understood with a simple 
crystal-field scheme of three levels at energies
 $\langle {\bar 2} |{\mathcal H}_{\rm CF}| {\bar 2} \rangle = \delta$, 
$\langle {\bar 3} |{\mathcal H}_{\rm CF}| {\bar 3} \rangle = \Delta$ with
$\Delta \gg \delta$ and zero. The ground state is the $| {\bar 1} \rangle$ state
with energy equal to zero. Since our interest is on the low-energy 
physics, we study  Eq.~\ref{Corr_13} with $m=1$ and $n=2$. In addition,
assuming $\delta$ negligible, 
\begin{eqnarray} 
\left (i \, \omega + \Gamma^{(12)}_{z,{\rm ph}} \right )Z_{z,12} 
=  \langle {\bar 1} | J_z| {\bar 2} \rangle ,
\label{Model_8}
\end{eqnarray}
where 
\begin{eqnarray} 
\Gamma^{(12)}_{z,{\rm ph}} &  = &  h_{\rm ph} \left (\Delta/\hbar  \right )
\left ( G_{13,31}^{\rm ph}  + G_{23,32}^{\rm ph} \right ). 
\label{Model_9}
\end{eqnarray}
The contributions to this equation come only from the two terms that 
satisfy the secular condition $\omega_{k \ell} =\omega_{m n}$ 
for $G^{\rm ph}_{k \ell, mn}$. The other terms oscillate with an angular frequency
$\omega_\Delta = \Delta/\hbar \simeq 10^{13} \, {\rm rad \, s}^{-1}$
for $\Delta/k_{\rm B} =100$~K,
and therefore average out to zero in the time scale of the measurements.
This simple model predicts a
quasi-elastic scattering. Assuming further $\beta \Delta \gg 1$
gives $\Gamma^{(12)}_{z,{\rm ph}} \propto \exp(-\beta \Delta )$
and therefore an activated behaviour for $\lambda_Z(T)$ is found. 
Components such as $Z_{z,31}$ are not relevant for the quasi-elastic 
response and $\lambda_Z$ since they are strongly inelastic. 
This argument is valid if the temperature is not too large so that  
$Z_{z,31}$ is a Lorentzian function of $\omega$ and so the contribution 
of $Z_{z,31}$ at $\omega=0$ is negligible. The relaxation
we are discussing is Orbach-like \cite{Orbach61}. It involves a 
real two-phonon scattering: one phonon induces a transition between 
$| {\bar 1} \rangle $ and
$|{\bar 3} \rangle$ and the other between $| {\bar 3} \rangle $ and
$|{\bar 2} \rangle$. It has been identified experimentally 
in $\mu$SR data \cite{Dalmas03}.
This scattering occurs only if phonons of the proper energy
are available. For a compound with a Debye phonon spectrum, this means
that  the excitation energy $\Delta$ cannot be larger than 
$k_{\rm B} \Theta_{\rm D}$ where $\Theta_{\rm D}$ is the Debye 
temperature. To complete our analysis of the high temperature regime, 
we note that $\Gamma^{(12)}_{z,{\rm ph}}$ is found to be proportional
to $T$ when $\beta \Delta \ll 1$. Therefore $\lambda_Z \propto 1/T$
when the thermal energy exceeds the crystal-field energy. 
Amusingly, the Korringa relaxation gives an identical temperature 
dependence. 

The same crystal-field scheme can be used to study the physics at low
temperature. The high-energy level is now neglected. Focusing 
again on the low-energy sector, $Z_{z,11}$ and $Z_{z,22}$ are the 
relevant components because $\delta$ is no more negligible. 
They obey a system of two coupled linear equations:
\begin{eqnarray} 
& & i \, \omega Z_{z,11} + 2 h_{\rm ph}(\delta/\hbar)
G^{\rm ph}_{12,21} \left ( Z_{z,11} - Z_{z,22}\right ) \cr
& = & \langle {\bar 1} | J_z| {\bar 1} \rangle -
\langle  J_z \rangle, \cr
& & i \, \omega Z_{z,22} - 2 g_{\rm ph}(\delta/\hbar)
G^{\rm ph}_{12,21} \left ( Z_{z,11} - Z_{z,22}\right ) \cr
& = & \langle {\bar 2} | J_z| {\bar 2} \rangle -
\langle  J_z \rangle.
\label{Model_10}
\end{eqnarray}
This system can be solved for $Z_{z,11}$ and $Z_{z,22}$ and 
$\Omega_{zz} (\omega)$ evaluated. Here we simply consider  
two limits. If $\beta \delta \gg 1$, $Z_{z,11}$ controls the
quasi-elastic linewidth. Since $h_{\rm ph}$ and 
$( \langle {\bar 1} | J_z| {\bar 1} \rangle  -
\langle J_z \rangle)$ vanish, the width of the quasi-elastic 
line vanishes also. Hence, the coupling of the strain field
to the crystal field does not induce a linewidth. 
In the limit $\beta \delta \ll 1$
, $Z_{z,11}$ and $Z_{z,22}$ have an equal weight in the expression 
for $\Omega_{zz}(\omega)$. Since they are of opposite sign,   
$\Omega_{zz}(\omega) =0$. Hence, in the high-temperature limit
of the two-level system the magneto-elastic coupling does not induce 
a linewidth. However, as our study of the three-level 
system has shown, the energy levels located at energy above $\delta$ 
may contribute to the relaxation.

In summary, in this section we have found that the magneto-elastic
coupling can effectively explain the activated behaviour of 
$\lambda_Z$ found at relatively high temperature for some 
insulators. The same coupling gives a vanishing linewidth
at low temperature. However, other mechanisms can contribute 
to it. Such a mechanism is unveiled in the next section.

\section{Spin-correlation for a regular lattice of paramagnetic ions}
\label{Lattice}

Here we study the effect of the magnetic interactions between 
the ions located in a regular crystal lattice. Hence, rather than 
${\mathcal H}_0 = {\mathcal H}_{\rm CF} + {\mathcal H}_{\rm L}$
with ${\mathcal H}_{\rm L} = {\mathcal H}_{\rm el}$ or
${\mathcal H}_{\rm L} = {\mathcal H}_{\rm ph}$ , we write
${\mathcal H}_0 = {\mathcal H}_{\rm CF} + {\mathcal H}_{\rm H} +
{\mathcal H}_{\rm L} $ where  
${\mathcal H}_{\rm H}$ describes the 
ion-ion magnetic interaction. We shall limit ourselves to the two-ion 
Heisenberg Hamiltonian, {\sl i.e.}
\begin{eqnarray}
{\mathcal H}_{\rm H} & = & -  \sum_{i,j} {\mathcal J (ij)}
{\bf J}_i \cdot{\bf J}_j,
\label{Lattice_0}
\end{eqnarray}
where ${\mathcal J (ij)} = 0$ when $i = j$. Introducing the following
relation in the previous equation,
\begin{eqnarray}
{\mathcal J (ij)} = {1 \over N} \sum_{\bf q} {\mathcal J_{{\bf q}}} 
\exp \left [i {\bf q} \left ({\bf i - j}  \right )  \right ],
\label{Lattice_0_1}
\end{eqnarray}
we derive
\begin{eqnarray}
{\mathcal H}_{\rm H} 
 =  - { 1 \over N}\sum_{{\bf q}} {\mathcal J_{ {\bf q}}} 
{\bf J}_{{\bf q}} \cdot{\bf J}_{-{\bf q}}.
\label{Lattice_0_2}
\end{eqnarray}
We have used the relation
\begin{eqnarray}
{\bf J}_{\bf q}  = \sum_i {\bf J}_i
\exp \left (-i{\bf q} \cdot {\bf i}  \right ).
\label{Lattice_0_3}
\end{eqnarray}
We recall that the magnetic ions are taken to sit on a Bravais lattice 
for simplicity.
As in Sec.~\ref{Lambda}, it is usually convenient to work with continuous 
rather than discrete wavevectors. This leads to 
\begin{eqnarray}
{\mathcal H}_{\rm H} 
 =  -  v \int {\mathcal J( {\bf q})} 
{\bf J}({\bf q}) \cdot{\bf J}(-{\bf q}) 
{{\rm d}^3 {\bf q} \over (2 \pi)^3},
\label{Lattice_0_4}
\end{eqnarray}
with 
\begin{eqnarray}
{\mathcal J} ({\bf q})
 & = & \sum_i {\mathcal J}(ij) 
\exp \left [-i {\bf q} \cdot \left ({\bf i - j} \right ) \right ].
\label{Lattice_0_5}
\end{eqnarray}

From now on we shall work with the reference frame 
$\{{\bf +}^\prime,{\bf -}^\prime,{\bf z} \}$ where the spin-correlation tensors 
are diagonal.
To simplify the notation and because  $\Lambda_{z}\left (\omega  \right )$ is 
simply $\Lambda_{zz}\left (\omega  \right )$ (see Sec.~\ref{Lambda}),
we shall only consider that correlation. Rather than $\Lambda_z(\omega)$ we have 
to evaluate the wavevector dependent symmetrised correlation function
\begin{eqnarray}
\Lambda_z({\bf q},\omega)& = & \int_{-\infty}^\infty 
\left \langle \{ J_z({\bf q},t) J_z({\bf -q},0) \}  \right \rangle 
\exp(i \omega t){\rm d} t, \cr
& & 
\label{Lattice_1}
\end{eqnarray}
where $2 \left \langle \{ J_z({\bf q},t) J_z({\bf -q},0) \}  \right \rangle 
= \left \langle J_z({\bf q},t) J_z({\bf -q},0)   \right \rangle + 
\left \langle J_z({\bf -q},0) J_z({\bf q},t)   \right \rangle $.

In general the available theories does not consider the correlation tensor but 
rather the generalized susceptibility tensor. The 
fluctuation-dissipation theorem provides a relation between the elements of the
two tensors, see for example  Ref.~[\onlinecite{Marshall71}]. Denoting the 
imaginary part of $\chi_z({\bf q},\omega)$ as 
$ {\mathcal Im} \{\chi_z({\bf q},\omega) \}$,
\begin{eqnarray}
\Lambda_z({\bf q},\omega) = 
{\hbar  V \over \mu_0 g^2 \mu_{\rm B}^2}
{\rm coth} \left ({\beta \hbar \omega  \over 2 }   \right )
{\mathcal Im} \{\chi_z({\bf q},\omega) \}.
\label{Lattice_2}
\end{eqnarray}
Therefore 
we need to compute $ {\mathcal Im} \{\chi_z({\bf q},\omega) \}$.

A practical theoretical method to describe the effect of the 
intersite correlations on the susceptibility 
is the RPA which leads to 
the following simple formula \cite{Jensen91}:
\begin{eqnarray}
{\bar {\bar \chi}}({\bf q},\omega) = \left \{ {\mathcal I} -  
{2 V {\mathcal J} ({\bf q})  \over \mu_0 g^2 \mu_{\rm B}^2}
{\bar {\bar \chi}}(\omega) \right \}^{-1}
{\bar {\bar \chi}}(\omega).
\label{Lattice_3}
\end{eqnarray}
$\{ A \}^{-1}$ stands for the inverse of tensor $A$.
${\bar {\bar \chi}}(\omega)$ is the susceptibility tensor for a magnetic ion 
in the lattice, the intersite interactions being switched off. 
The elements of the tensors ${\bar {\bar \chi}}(\omega)$ and 
$\bar {\bar \Lambda}(\omega)$ are related. Using again the 
fluctuation-dissipation theorem, 
\begin{eqnarray}
{\mathcal Im} \{\chi_z(\omega)\} =
{\mu_0 g^2 \mu_{\rm B}^2  \over  \hbar V}
{\rm tanh} \left ({\beta \hbar \omega \over 2}  \right )
\Lambda_z( \omega).
\label{Lattice_5}
\end{eqnarray}
The real and imaginary parts of $\chi_z(\omega)$ are linked through 
the Kramers-Kronig theorem:
\begin{eqnarray}
{\mathcal Re} \{\chi_z(\omega) \}& = & 
-{1 \over \pi} {\mathcal P}\int_{-\infty}^\infty    
{{\mathcal Im}\{\chi_z(u) \}
\over \omega - u}{\rm d} u.
\label{Lattice_6}
\end{eqnarray}
${\mathcal P}$ stands for the principal part of the integral.
Thus, once tensor elements such as $\Lambda_z( \omega)$ are known, 
the elements of ${\bar {\bar \chi}}(\omega)$ can be computed using 
Eqs.~\ref{Lattice_5} and \ref{Lattice_6}. Then, 
${\mathcal Im} \{\chi_z({\bf q},\omega)\}$ can be determined from
Eq.~\ref{Lattice_3}. Finally $\Lambda_z (\bf q{},\omega)$ is obtained 
recalling Eq.~\ref{Lattice_2}.
Within our conventions ${\bar {\bar \chi}}({\bf q},\omega)$ and
${\bar {\bar \chi}}(\omega)$ are dimensionless. 

Since the tensors are diagonal in the reference frame 
$\{{\bf +}^\prime,{\bf -}^\prime,{\bf z} \}$, Eq.~\ref{Lattice_3} can be solved easily 
for ${\bar {\bar \chi}}({\bf q},\omega)$. Combined with Eq.~\ref{Lattice_2},
\begin{widetext}
\begin{eqnarray}
\Lambda_z (\bf q{},\omega) & = &  
{ \hbar V \over \mu_0 g^2 \mu_{\rm B}^2} 
{\rm coth} \left ({\beta \hbar \omega \over 2}  \right )   
{\mathcal Im} \left \{ {\chi_z(\omega) \over 
1 - {2 V  {\mathcal J} ({\bf q})  \over \mu_0 g^2 \mu_{\rm B}^2}
\chi_z(\omega)} \right \} \cr
& = & {\Lambda_z(\omega)  \over 
\left [1 + {2 {\mathcal J} ({\bf q})   \over \pi \hbar}
{\mathcal P}\int_{-\infty}^\infty {\Lambda_z(u) \over \omega - u}
{\rm tanh} \left ({\beta \hbar u \over 2}  \right ) {\rm d} u \right ]^2 + 
\left [ { 2 {\mathcal J} ({\bf q}) \over \hbar}
{\rm tanh} \left ({\beta \hbar \omega \over 2}  \right )
 \Lambda_z(\omega) \right ]^2}.
\label{Lattice_7}
\end{eqnarray}
\end{widetext}

Therefore, if $\Lambda_z (\omega)$ and ${\mathcal J} ({\bf q})$ are known,
it is possible to compute numerically $\Lambda_z (\bf q{},\omega)$
and $\lambda_Z$ for a given muon site using the material of 
Sec.~\ref{Lambda}. We can proceed further analytically for the elastic
case, {\sl i.e.} when $\omega =0$. For definitiveness, we 
assume $\Lambda_z(\omega)$ to be Lorenzian, {\sl i.e.}
\begin{eqnarray}
\Lambda_z (\omega) & = &  
 {2 {\langle J_z^2 \rangle}  \Gamma_z  \over 
\omega^2 + \Gamma^2_z}
\label{Model_11}
\end{eqnarray}
We derive
\begin{eqnarray}
& & \Lambda_z ({\bf q},\omega=0) \cr & = & 
{\Lambda_z (\omega=0) \over 
\left [1 - 
{4 {\langle J_z^2 \rangle}  {\mathcal J} ({\bf q})
\over \hbar \Gamma_z} {1 \over \pi}
\int_{-\infty}^\infty {1 \over  x(x^2 + 1)}
\tanh \left ({\beta \hbar \Gamma_z \over 2}x  \right ) {\rm d} x
\right ]^2}
\cr & = &  
 {2 {\langle J_z^2 \rangle} \over 
\Gamma_z 
\left [1 - 
{4 {\langle J_z^2 \rangle}  {\mathcal J} ({\bf q})
\over \hbar \Gamma_z} {1 \over \pi}
\int_{-\infty}^\infty {1 \over  x(x^2 + 1)}
\tanh \left ({\beta \hbar \Gamma_z \over 2}x  \right ) {\rm d} x
\right ]^2}. \cr
& & 
\label{Model_12}
\end{eqnarray}

To understand this expression we consider the consequences of a magnetic phase 
transition. That transition is characterized, among other properties, by   
its critical temperature $T_c$ and a wavevector ${\bf q}_c$ 
which sets the repetition length of the magnetic structure for $T < T_c$. 
At $T_c$ the generalized susceptibility diverges for 
${\bf q} = {\bf q}_c$, and so does $\Lambda_z({\bf q}_c,\omega=0)$. From 
the previous equation
\begin{eqnarray}
{\hbar \over  {\mathcal J} ({\bf q}_c)} = 
{{4 {\langle J_z^2 \rangle}} \over \Gamma_z} { 1 \over \pi}
\int_{-\infty}^\infty {1 \over  x(x^2 + 1)}
\tanh \left ({\hbar \Gamma_z \over 2 k_{\rm B} T_c}x  \right ) {\rm d} x. \cr
& & 
\label{Model_13}
\end{eqnarray}
Let us study 
$I(\alpha) = \int_{-\infty}^\infty {1 \over  x(x^2 + 1)} \tanh (\alpha x) {\rm d} x$. 
We evaluate numerically $I(1) = 2.08$. This is clearly 
smaller than $\int_{-\infty}^\infty {1 \over  x^2 + 1}  {\rm d} x = \pi$. However,
smaller is $\alpha$ better is the approximation of linearising the function
$ \tanh (\alpha x)$. In our case $\alpha = (\hbar \Gamma_z )/ (2 k_{\rm B} T_c)$. 
Physically, we expect $\hbar \Gamma_z \simeq k_{\rm B} T_c$ for a conventional phase 
transition. Assuming the linearisation to be valid,  we get
\begin{eqnarray}
{\mathcal J} ({\bf q}_c) =  {{k_{\rm B} T_c \over 2  {\langle J_z^2 \rangle}}}.
\label{Model_14}
\end{eqnarray}
This is consistent with the mean-field result.

We known that $\lambda_Z$ in zero field is given as a sum over the Brillouin zone of 
$\Lambda_z({\bf q}, \omega=0)$. Approaching the critical point, the sum is expected 
to be dominated by $\Lambda_z({\bf q}_c, \omega=0)$ since that correlation diverges
at $T_c$. How does it diverge? According to our RPA result of 
Eq.~\ref{Model_12}, upon linearising the $\tanh$ function, we derive
\begin{eqnarray}
\Lambda_{z}({\bf q}_c,\omega=0) & = &
\Lambda_{z}(\omega= 0) \left ({T \over
T - T_c }\right )^2.
\label{Lattice_12_1}
\end{eqnarray}
While Eq.~\ref{Lattice_12_1} is obviously not valid in the critical regime
since the RPA is a kind of mean-field approximation, it should provide 
a good description deep in the paramagnetic regime. Note that 
$\Lambda_{z}({\bf q}_c,\omega=0)$ reaches a finite temperature independent
value at high temperature. 

Now we study $\Lambda_z({\bf q},\omega=0)$ in the limit $T \gg T_c$
($T_c = T_{\rm C}$ for a ferromagnet and  $T_c = T_{\rm N}$ for an antiferromagnet).
We find:
\begin{eqnarray}
\Lambda_{z}({\bf q},\omega=0) & = &
\Lambda_{z}(\omega= 0) \left (1 + 2
{ {\mathcal J} ({\bf q}) \over 
{\mathcal J} ({\bf q}_c) } 
{ T_c \over T}\right ).
\label{Lattice_12}
\end{eqnarray}
Because ${\mathcal J} ({\bf q}_c) > 0$, 
the sign of the thermal correction at high temperature is given by the sign of 
${\mathcal J} ({\bf q})$. Referring to Eq.~\ref{Lattice_0_4},
positive and negative ${\mathcal J} ({\bf q})$ values correspond
to ferromagnetic and  antiferromagnetic interactions at 
wavevector ${\bf q}$, respectively. This means that a ferromagnetic
(antiferromagnetic) interaction increases (decreases) the correlation
function $\Lambda_{z}({\bf q},\omega=0)$. This result was derived a 
long time ago, but the wavevector dependence of the interaction was 
overlooked \cite{Silbernagel68,Hartmann86}. The $1/T$ dependence
of $\Lambda_{z}({\bf q},\omega=0)$ is a typical pair correlation 
effect at high temperature.  

The result of the RPA approximation is quite simple, in
particular for $\Lambda_z (\bf q{},\omega=0)$. When $ \Lambda_z (\omega = 0)$
is determined, $\Lambda_z (\bf q{},\omega=0)$ can be evaluated by
dividing  $ \Lambda_z (\omega = 0)$ by a proper denominator.
However, in Sec.~\ref{Model} we have found   
$ \Lambda_z (\omega = 0)$ to vanish if the thermal energy is much larger
than the crystal-field energy $E_{\rm CF}$ when the magneto-elastic coupling
drives the crystal-field relaxation. In this case $\lambda_Z$
should drop as the compound is warmed up far above $E_{\rm CF}/k_{\rm B}$
since $\Lambda_z (\bf q{},\omega=0)$ becomes negligible. This is inconsistent 
with the experimental finding for the garnet Yb$_3$Ga$_5$O$_{12}$ \cite{Dalmas03}.
To understand the origin of this problem, we need first to specify the 
Yb$^{3+}$ level scheme. Because Yb$^{3+}$ is a Kramers ion, the states 
are at least doubly degenerate. The mean distance between 
the doublet ground state and the three closely spaced excited 
doublets is $\Delta/k_{\rm B} \simeq 850$ K \cite{Buchanan67}. 
Since the ground state is a doublet, $E_{\rm CF} = 0$. In addition,
$\beta \Delta \gg 1$. Hence, from the RPA approximation we expect an activated 
behaviour for $\lambda_Z(T)$ outside the critical regime ($T_{\rm N} =0.05$ K).
But experimentally, from about $0.4$ up to 
$100$ K, $\lambda_Z$ is temperature independent. It displays 
an activated behaviour only above $\sim 100$ K.

The breakdown of the RPA theory for Yb$_3$Ga$_5$O$_{12}$ could have been 
anticipated. In fact, the compound behaves as a lattice of effective spins one-half 
coupled by the Heisenberg interaction
up to $\sim 100$~K.
This situation cannot be described by the RPA approximation we use because it assumes 
a non-zero single ion susceptibility \cite{Jensen91}. Only, if the temperature is 
increased to be such that the thermal energy is an appreciable fraction of the crystal
field excitation energy, does 
${\mathcal H}_{\rm CF}$ starts to matter and the material given in
Secs.~\ref{Correlation} and \ref{relaxation} becomes relevant. The Orbach
relaxation then becomes visible.

To model the quasi-elastic response and $\lambda_Z$ for
Yb$_3$Ga$_5$O$_{12}$ at intermediate temperature, 
a finite expression of the self-correlation 
function $\Lambda_z(\omega)$ for an isotropic magnetic material in the 
limit $\omega \rightarrow 0$ is needed. 
That function is given by
the sum over the Brillouin zone of $\Lambda_z({\bf q},\omega)$. According to the 
fluctuation-dissipation theorem, $\Lambda_z({\bf q},\omega)$ is proportional to 
the product of the static wavevector-dependent susceptibitity, 
$\chi_{z}({\bf q})$, and the spectral-weight function, $F_z ({\bf q},\omega)$.
From the material gathered by Lovesey \cite{Lovesey86a}, we could write
a mean-field expression for $\chi_{z}({\bf q})$ and $F_z ({\bf q},\omega)$
in the Gaussian approximation. Defining $\Gamma_{z, {\rm exc}}$ such that 
$\Lambda_z(\omega=0) = 2 \langle J^2_z \rangle/\Gamma_{z, {\rm exc}}$,
we could identify $\Gamma_{z, {\rm exc}}$ by setting $\Lambda_z(\omega=0)$
equal to the sum over the Brillouin zone of $\Lambda_z({\bf q},\omega=0)$.
However, the formulae for $\chi_{z}({\bf q})$ and $F_z ({\bf q},\omega)$
are approximate and the sum can only be done numerically. Here we propose
a much simpler method to estimate $\Gamma_{z, {\rm exc}}$. We suppose the
spectral function of the self-correlation to be Gaussian and use the 
expression of the second moment for paramagnets given in 
Ref.~[{\onlinecite{Lovesey86a}}]. We get:  
\begin{eqnarray}
\Gamma_{z,{\rm exc}} = \sqrt {16 {\mathcal J}^2 z J(J+1)\over 3 \pi \hbar^2}.
\label{Lattice_13}
\end{eqnarray}
Interestingly, $\Gamma_{z,{\rm exc}}$ depends on the 
two-ion exchange interaction ${\mathcal J}$ (see Eq.~\ref{Lattice_0_5};
${\mathcal J} = {\mathcal J}(ij)$) 
that we have limited to the $z$ nearest neighbour ions. Therefore
it reflects the environment of the ions under
study.  $\Gamma_{z,{\rm exc}}$ is temperature
independent in the high-temperature limit we take and for the conventional magnets
we consider, {\sl i.e.} we do not discuss complicated compounds such as the 
frustrated ones.

The inclusion of the phonon-induced relaxation is easily done.
Referring to Eq.~\ref{Model_11} for the definition of
$\Gamma_z$,
\begin{eqnarray}
\Gamma_z = \Gamma_{z,{\rm exc}} + \Gamma_{z,{\rm ph}}.
\label{Lattice_14}
\end{eqnarray}
In the same way, for the electronic relaxation we have
\begin{eqnarray}
\Gamma_z = \Gamma_{z,{\rm exc}} + \Gamma_{z,{\rm el}}.
\label{Lattice_15}
\end{eqnarray}
$ \Gamma_{z,{\rm el}}$ and $\Gamma_{z,{\rm ph}}$ are obviously 
obtained from the study of $\Omega_{zz}(\omega)$. 
Eq.~\ref{Lattice_14} has been proven to provide a good fit of 
$\lambda_Z(T)$ observed for Yb$_3$Ga$_5$O$_{12}$ \cite{Dalmas03},
but only for $T > 0.4$ K. The increase of $\lambda_Z$ 
for $T_c < T < 0.4$ K arises from the intersite correlations. 
The RPA provides a model for these correlations, but outside of the critical 
regime.

In conclusion, the analysis of the neutron quasi-elastic linewidth  and $\mu$SR 
spin-lattice relaxation rate 
should be done, at least in a first approximation, with Eq.~\ref{Lattice_7} or 
Eq.~\ref{Model_12} and 
using for the single ion linewidth the expression given at
Eq.~\ref{Lattice_14} or Eq.~\ref{Lattice_15}.

\section{Discussion and conclusion}
\label{Discussion}

We have studied the symmetrised correlation functions as measured
by neutron scattering and the $\mu$SR spin-lattice relaxation rate. 
We have focused our work on the paramagnetic state 
of magnetic materials made of a regular lattice of rare-earth ions. 
The relaxation of the crystal-field levels has been assumed to be 
driven by the conduction electrons or the lattice vibrations. 
Mathematically, we have used the iterative Laplace transform method 
due to P. M. Richards. The interaction between the ions has been
modelled with the random-phase approximation. A phenomelogical 
modification of this approximation has been proposed to cure its 
breakdown which comes into light when the crystal-field levels are 
not closely packed. 
Thanks to the point symmetry at the rare-earth site, for many 
compounds of interest, we have shown that at most two 
symmetrised correlation functions are relevant.

Through the study of simple crystal-field models, the Orbach relaxation 
mechanism has been uncovered for the linewidth and $\lambda_Z$. We 
have argued for a single ion that the linewidth should vanish at low 
temperature if the relaxation is driven by the lattice vibrations. 
However, physically this limit cannot be reached because of the on-site 
fluctuations described by $\Gamma_{z,{\rm exc}}$ in Sec.~\ref{Lattice},
which set a lower bound on the linewidth and therefore an upper 
bound on $\lambda_Z$. 

As clearly seen from its derivation, the Richard's system of linear 
equations is obtained if the Hamiltonian describing the interaction
of the lattice field with the crystal field can be factorised into 
these two fields. This sets limits on the relaxation mechanisms which 
can be handled. For example, the Raman mechanism cannot be included. 
It involves two virtual phonons and it is described by treating the 
interaction Hamiltonian in second order \cite{Orbach61}. It leads 
to a linewidth $\Gamma_{\rm ph} \propto T^{-n}$ with 
a large value for the exponent $n$, typically $n=5$ or $n=7$. 
According to Electron Paramagnetic Resonance data, this mechanism 
can appear only at low temperature \cite{epr66}.

The modulation of the crystal field has been expanded in powers 
of the strain and we have kept the linear term. We could have 
gone a step further, keeping the term quadratic in strain. This 
would have introduced at least one new free parameter. Unless
really justified for a particular physical problem, we do not 
believe it is worthwhile.

Although our interest in this paper has been on the paramagnetic 
phase of a magnetic material, the method we have discussed 
can be extended to its ordered magnetic state. One has to account for 
the molecular field acting on the rare-earth ion by adding a Zeeman 
term to ${\mathcal H}_{\rm CF}$ as done in Ref.~[\onlinecite{Purwins73}].  
The RPA models the magnetic fluctuations due to the two-ion interaction.

In conclusion, we have proposed a framework to analyse the linewidth
as measured by neutron scattering and the $\mu$SR relaxation rate for 
compounds with rare-earth sublattices. We take into account the electronic 
and phonon-induced relaxation of the crystal-field levels. We have proposed a 
phenomelogical modification of the result of random-phase approximation
which appears for some particular crystal-field level schemes.

I thank Pierre Dalmas de R\'eotier for many discussions and his checking of
equations.

\bibliography{reference}

\end{document}